\documentclass[prd,twocolumn,eqsecnum,showpacs,nofootinbib]{revtex4}
\usepackage{bm}
\usepackage{amssymb}
\usepackage{graphicx}
\usepackage{epstopdf}

\begin{document}
\title{Self--force gravitational waveforms for extreme and intermediate mass ratio inspirals. II: Importance of the second--order dissipative effect}
\author{Lior M.~Burko$^{1,2}$ and Gaurav Khanna$^3$}
\affiliation{$^1$ Department of Physics, Chemistry, and Mathematics, Alabama A\&M University, Normal, Alabama 35762 \\
$^2$ Theiss Research, La Jolla, California 92037\\
$^3$ Department of Physics, University of Massachusetts, Dartmouth, Massachusetts  02747}
\date{April 18, 2013}
\begin{abstract} 

We consider the importance of the second--order dissipative self force for gravitational wave dephasing for an extreme  or intermediate mass ratio system moving along a quasi-circular Schwarzschild orbit. For the first--order self force we use the fully relativistic force in the Lorenz gauge for eternally circular geodesics.  
The second--order self force is modeled by its 3.5 post Newtonian counterpart. We evolve the system using the osculating orbits method, and obtain the gravitational waveforms, whose phase includes all the terms ---within our approximation (and using the self force along circular geodesics)--- that are independent of the system's mass ratio. 
The partial dephasing due to the second--order dissipative self force is substantially smaller than that of the first--order conservative self force, although they are both at the same order in the mass ratio. 

\end{abstract}
\maketitle

\section{Introduction and summary}

When a small compact object moves in the spacetime of a much more massive black hole it accelerates and emits gravitational waves. The cause of the acceleration of the smaller object $\mu$ may be viewed as the object's interaction with itself in the external field of the more massive black hole $M$, namely the object's self force (henceforth the SF). The SF causes a correction to geodesic motio in the mass ratio of the system $\varepsilon :=\mu /M$. The force on $\mu$ vanishes for geodesic motion. Its leading term in an expansion in $\varepsilon$, that is the force at $O(\varepsilon^2)$, has been studied in detail\footnote{For enumeration purposes it is useful to think in terms of the SF per unit mass. Then the leading order term is at $O(\epsilon)$ and the next term at $O(\varepsilon^2)$, which justifies the names first order and second order SF, respectively. 
As we refer here to the SF, not the SF per unit mass,  the first order SF is at $O(\varepsilon^2$) and the second order SF is at $O(\varepsilon^3)$.} \cite{poisson,barack-2009}. For more detail, see the Introduction to Ref.~\cite{lackeos_burko_12} (hereafter Paper I) that will serve as an introduction to the present Paper too. 

It has been realized, however, that the dephasing of gravitational waveforms includes important contributions also from the SF at $O(\varepsilon^3)$ \cite{burko-2003}. We illustrate this point for quasi-circular equatorial orbits of $\mu$ around a Schwarzschild black hole $M$: The SF in this case can easily be separated between its dissipative and conservative pieces; the dissipative piece is made of the temporal and azimuthal components of the self force, while the conservative piece is described by the radial component thereof. We schematically expand the phase of the gravitational waveforms in powers of $\varepsilon$ as follows \cite{flanagan-hinderer}:
\begin{equation}\label{pert}
\Phi=\varepsilon^{-1}\left[\,\Phi^{(0)}+\varepsilon \, \Phi^{(1)}+O(\varepsilon^2)\right]\, .
\end{equation}
The dissipative SF at $O(\varepsilon^2)$ causes the phase of the emitted gravitational wave to evolve at $O(\varepsilon^{-1})$, that is it determines $\Phi^{(0)}$. In essence, that is the same waveform as obtained when the orbital evolution is computed used balance arguments. The dephasing of the waveform at $O(\varepsilon^0)$, given by $\Phi^{(1)}$, has two contributions\footnote{In our approach there is a third contribution: as we use in practice the SF along an eternal circular geodesic, the difference between this SF and the true SF (that includes the effects of deviations of the orbit from a circular geodesic) also introduces a contribution to $\Phi^{(1)}$. This effect is neglected in this Paper: see below.}: both the conservative piece of the SF at $O(\varepsilon^2)$ and the dissipative piece of the SF at $O(\varepsilon^3)$ contribute to the dephasing of the waveform at the same order, a dephasing which is independent of the system's mass ratio.

It was shown in Paper I that the dephasing $\Phi^{(1)}$ at $O(\varepsilon^0)$ of the gravitational waveform due to the first--order conservative piece of the SF at $O(\varepsilon^2)$ is important for gravitational wave detection and astronomy. In this Paper we consider the other source of dephasing of the waveform at the same order, that is at $O(\varepsilon^0)$, specifically the second--order dissipative piece of the SF at $O(\varepsilon^3)$, and study its relative importance to the dephasing. 

The SF we use in practice is the SF acting on a particle moving for all eternity along a circular geodesic. As the particle does not move in actuality along an eternal circular orbit, but rather along a slowly decaying quasi-circular orbit, the actual SF differs from the SF obtained for circular geodesic by terms which are second order at $O(\varepsilon^3)$: the true SF is obtained from integration over the half infinite world line of the particle. Specifically, when one computes the leading order dissipative SF for the inspiralling world line, it would differ from its circular geodesic counterpart by terms at  $O(\varepsilon^3)$. We neglect such  $O(\varepsilon^3)$ corrections to the SF. Our computational method is the method of osculating orbits \cite{pound-poisson}; specifically, we use the fact that at each point along the orbit the orbit locally osculates to a geodesic, and calculate the parameters of the geodesic that slowly evolve along the orbit. This approach may prevent the $O(\varepsilon^3)$ error  from growing uncontrollably along the integration of the orbit. Combining the osculating orbit equations with the force as calculated on geodesics has been used before by Warburton {\em et al} \cite{warburton} and by Lackeos and Burko \cite{lackeos_burko_12}. Here we continue with the same approach, but carry it to consider the relative importance of the various partial dephasings at $O(\varepsilon^0)$. 

We propose the viewpoint that the various terms that contribute to dephasing at $O(\varepsilon^0)$ do so independently of each other at leading order. Specifically, the partial dephasing due to the first order conservative effect is independent to leading order of the partial dephasing due to the second order dissipative effect, or the correction to the first order dissipative effect that results from the fact that the true orbit is not an eternally circular geodesic. We study the relative importance of the first two effects for quasi--circular Schwarzschild orbits, and show that partial dephasing due to the second effect is significantly smaller than the partial dephasing due to the first. This conclusion is independent of whether the partial dephasing due to the last effect is also negligible or not.

%Therefore, the consistency of our approach allows us to consider the SF at each point along the orbit as that of an eternal circular geodesic, so that in fact we include {\it all} $O(\varepsilon^0)$ terms to the phase $\Phi$ in our osculating orbits approach. 

Despite impressive progress in the understanding of the second-order SF (at $O(\varepsilon^3)$) \cite{gralla-2012,pound-2012a,pound-2012b}, it is as yet unavailable for us for practical use. We may however understand its significance qualitatively by mimicking it with its post Newtonian (PN) counterpart. (Our analysis can readily be adapted to the fully relativistic second order SF when the latter becomes available.) However, our conclusions based on a hybrid model in which the first order SF is fully relativistic (specifically we use the Barack--Sago SF \cite{barack-sago}) and the second order is approximated by the post Newtonian expression will remain qualitatively unchanged when the latter is replaced by its fully relativistic counterpart. We emphasize that our results are quantitatively accurate only in the far field, but close to the innermost stable circular orbit (ISCO) we lose accuracy, because of our use of the post Newtonian expression for the second order SF. Specifically, the second order dissipative SF is known to 3.5 PN order, that is to $O(v^7)$ \cite{blanchet}. It is listed explicitly in Appendix \ref{PN}. The first order dissipative self force at 3.5 PN overestimates the fully relativistic Barack--Sago self force by not more than $10\%$ approaching the ISCO (and significantly less away from the ISCO). It is reasonable to expect that a similar accuracy is provided also by the second order dissipative force. Indeed, Isoyama {\em et al}  \cite{isoyama_2012} showed that the dephasing due to the $O(\varepsilon^3)$ SF for quasi-circular extreme mass--ratio inspirals (EMRIs) may be well captured by the 3 PN term. 

Unlike its scalar field counterpart, the gravitational SF is gauge dependent. Specifically, one can find a gauge in which the SF vanishes (a problem that would be equivalent to computing the SF driven orbital evolution), or gauges that satisfy certain algebraic or differential relations. The SF that we use in practice at first order, namely the Barack--Sago SF \cite{barack-sago}, is given in the Lorenz gauge. Moreover, the orbit itself is also a gauge dependent quantity. Fixing the gauge to the Lorenz (or harmonic) gauge, we focus attention on gauge invariant quantities (``observables"), such as the waveforms and the plot of $u^t$,  the temporal component of the four-velocity (``gravitational redshift,"``helical Killing vector of the perturbed spacetime") as a function of the angular frequency $\Omega$. 

Although the waveform and other gauge invariant quantities are exactly gauge invariant when treated self--consistently, we only have an approximation thereof because of our neglecting of the correction to the first order SF because of the evolution of the orbit. Therefore we propose that our waveform is not exactly gauge invariant, but only ``approximately gauge invariant" within a class of gauges for which the gauge-change in the metric perturbations remains $O(\varepsilon)$ over a radiation reaction timescale; in this class of gauges the metric perturbations respect the approximate helical symmetry of the orbit over the entire inspiral.

An important question is whether we are allowed to take the first order SF in one gauge (specifically the Lorenz gauge) and the second order SF in another gauge (specifically the harmonic gauge of PN theory). In general such hybridization is likely to fail. However, in our case the two gauges in question -- specifically the Lorenz and the harmonic gauges -- share important mathematical properties and singular behavior approaching the point particle. Indeed, to leading order in perturbation theory over Minkowski spacetime the Lorenz gauge and the harmonic gauge are exactly equivalent \cite{poisson-will}. It appears plausible that over curved background, waveforms constructed from these two gauges would be equivalent up to some post Newtonian order. In practice, we find at first order the numerical values of the Lorenz gauge SF and the harmonic gauge SF to agree to the number of significant figures relevant to our analysis. We therefore have a consistent use of gauge, specifically the harmonic gauge: Even though the fully--relativistic first order SF is obtained in the Lorenz gauge, it is in practice equivalent to the first order SF in the harmonic gauge; our second order SF is already in the harmonic gauge, so that the combined SF is consistently in the harmonic gauge. We further raise the question of whether for the class of orbits studied here this equivalency of the gauges goes beyond the leading order in perturbation theory up to some PN order. If that were the case, then our SF is also fully consistent in the Lorenz gauge.

Working in the regular Schwarzschild coordinates, we first evolve the orbit and find the gravitational waveforms when the orbital evolution is driven by the first order SF (similarly to the analysis in Paper I), and then we add the contribution of the second order dissipative SF. We are mostly interested in the relative dephasing of the two waveforms ---that is, in the partial dephasing due to the second--order dissipative SF--- and specifically in how large this dephasing is compared with the relative dephasing between the energy balance waveform and the waveform obtained with the inclusion of the first order conservative SF (omitting the second order SF, as was done in Paper I) ---the partial dephasing due to the first--order conservative SF. 

We find that in the parameter range we search, specifically the mass ratios $10^{-5}$--$10^{-2}$ the contribution to the dephasing because of the second order dissipative effect is much smaller than the contribution of the first order conservative effect, although they are both at the same order in that mass ratio (they are both independent of the mass ratio). For the orbits we describe in detail, specifically quasi-circular Schwarzschild orbits that start at $r_0=8M$ and evolve down to the ISCO, the dephasing due to the second order dissipative effect amounts to $8\%$ (with a negative sign) of the contribution of the first order conservative effect. In the gauge invariant plot showing $u^t$ as a function of the angular frequency $\Omega$, the effect of the second order dissipative self force is even smaller, and is manifested by a slower motion of the data point representing the system along the curve as the system is evolving. The smallness of the effect may suggest that it is unimportant for gravitational wave detection, although it may be importance for accurate gravitational wave astronomy, e.g., for precise parameter estimation. Of course the effect could be of the utmost importance for a chaotic system. Although chaos has been shown to exist for certain Kerr orbits with spinning particles, such orbits are not likely to be relevant for EMRI gravitational wave astronomy \cite{hartl}. 

We emphasize that even though we include in our waveforms the second--order dissipative SF in addition to the first--order conservative SF, so that all sources of dephasing at order $O(\varepsilon^0)$ are included (neglecting the contribution of the error in the SF because of its difference in actuality from the SF along an eternal circular geodesic), the resulting orbits and corresponding waveforms are not exactly the same as their self--consistent counterparts, except perhaps in the adiabatic limit. To obtain true self--consistent orbits  and corresponding waveforms one needs to simultaneously obtain the instantaneous solution of the coupled SF integrated equations of motion and the perturbation equations. Self consistent evolutions have been obtained for scalar field SF \cite{diener-2012}, yet not for the gravitational SF. Althought self--consistent evolutions have many advantages, we use instead the fully relativistic SF that was obtained for exact circular orbits \cite{barack-sago} within the osculating orbits approach. In the adiabatic limit the error introduced by the latter method should become negligible. Our waveforms may serve to test self--consistent waveforms when the latter become available, and provide a computationally effective method to compute practical waveforms if the errors included in our approach are found to be negligible ---at least for some classes of orbits. 

The organization of this Paper is as follows: In Section \ref{numerical} we describe the numerical method that allows us to evolve the orbit and to extract the corresponding gravitational waveforms with high accuracy and computational efficiency. In Section \ref{simulations} we describe our simulations and find the waveforms, and finally in Section \ref{dephasing} we analyze the results and find the dephasing due to the effect of interest, and in Section \ref{results} we describe the gauge invariant plot. 

\section{The numerical method}\label{numerical}

In our time-domain approach towards numerically modeling an EMRI system, there are two distinct computations that need to be performed very accurately. First, is the computation of the orbital trajectory of the small object in the spacetime of the large black hole including the SF effects under consideration in this work. This has been performed using the osculating orbits method that we describe in Paper I. Since we showed in Paper I that this method is reliable, as the produced orbits are consistent with those obtained with the independent direct integration method, we use here the osculating orbits method exclusively. As shown in Paper I, our osculating orbits code converges with 5$^{\rm th}$ order. Special care was taken to make certain of the accuracy of the trajectory throughout its rather long duration (on the order of 10,000 cycles for $\varepsilon=10^{-5}$) in this work. 

Second, we use this trajectory generated in the previous step as input to our time-domain Teukolsky EMRI code. This code, that numerically solves the inhomogeneous Teukolsky equation with a particle-source, has been used for studying EMRI related problems for over a decade. Essentially, the Teukolsky EMRI code is linear, hyperbolic (2+1)D PDE solver using a time-explicit, Lax-Wendroff, 2nd-order finite-difference numerical evolution scheme. The particle-source term on the right-hand-side of the equation requires some care for a numerical implementation. Details on the capabilities, accuracy, convergence and performance of the code appear at multiple places in the literature (see Ref.~\cite{prx2011} for a recent review and \cite{zenginoglu} for mathematical background). 

In order to be able to efficiently perform the type of long evolutions (on the order of a {\em billion} time steps) being presented in this work, we used an advanced high-performance version of the Teukolsky EMRI code, that is designed to execute on a large parallel supercomputer accelerated by general-purpose graphics processing units (GPUs). Details on the OpenCL-based parallel implementation and careful measurements of gains in overall code performance can be found in Ref.~\cite{xsede12} (and references therein). This code also has an added compactified hyperboloidal layer to the outer portion of our computational domain~\cite{prx2011}. This advancement allows us to map null infinity to the computational grid and thus, we are able to extract gravitational waveforms directly at null infinity. These technical advancements have helped towards improving the performance and accuracy of our time-domain Teukolsky EMRI code by several orders-of-magnitude over previous versions. This code currently performs at a level of accuracy on the scale of a {\em hundreth of a percent} while still maintaining a very high degree of computational efficiency~\cite{xsede12}\footnote{The longest evolution we have performed to date is for a binary system of $\varepsilon=10^{-6}$ with the central black hole carrying a spin of $a/M=0.9$. The inspiral trajectory starts out at $r=5M$ and lasts nearly $1.6 \times 10^7 M$ in duration, or nearly $300,000$ orbital cycles before it plunges into the central hole. That 2+1D Teukolsky equation based computation was performed using 600 processors-cores accelerated by 150 GPGPUs in just under 24 hours.}.

\section{The perturbative waveforms}\label{simulations}

\subsection{The perturbative expansion}

%The perturbative approach that underlies Eq.~(\ref{pert}) assumes that one may expand the force acting on the small object $\mu$ in powers of the mass ratio $\varepsilon$, that the resulting orbital evolution can be similarly expanded, and that these expansion properties carry over to the corresponding waveforms. For $\varepsilon\to 0$ this is surely the case, but any physical system has a finite value for $\varepsilon$. 

In Paper I we considered $10^{-3}\lesssim \varepsilon \lesssim 10^{-2}$ for practical purposes: the number of orbits and the resulting evolution time and number of numerical steps needed grow with $\varepsilon^{-1}$. While certain astrophysical systems of interest indeed have mass ratios in the range studied in Paper I, we would like to study also mass ratios at the order of $\varepsilon\sim 10^{-5}$, and in addition also quantify the mass ratio range for which our perturbative expansion of the orbit and the corresponding waveform is appropriate. The improvements in the computational technique allow us to do that. 

The first restriction, as discussed in Paper I, is that the adiabatic condition holds, or that 
$\delta\gg\varepsilon^{1/2}$, where $\delta:=p-6-2\epsilon$, $p$ is the orbit's semilatus rectum, and $\epsilon$ is the orbital eccentricity \cite{cutler}.  This condition does not guarantee, however, that the corrections to the orbital motion in powers of $\varepsilon$ drop sufficiently fast. 

\begin{figure}
 \includegraphics[width=7.5cm]{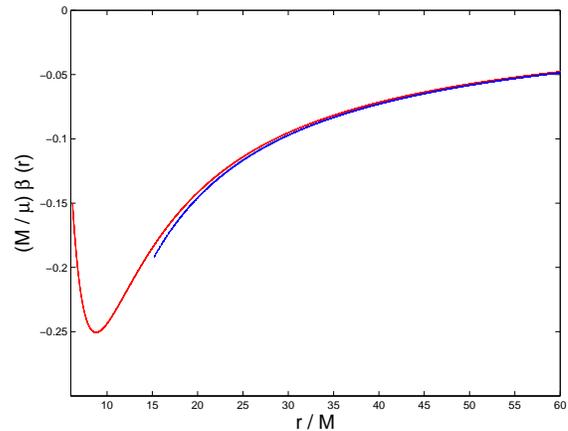}
\caption{The quantity $\varepsilon^{-1}\beta$ as a function of $r/M$. The long (red) curve is the curve based on the fully relativistic Barack--Sago results for $f^t_{(1)}$, and the shorter dotted (blue) curve represents the first  term in the post Newtonian expansion.}
\label{expand}
\end{figure}

The expansion of the SF in powers of $\varepsilon$, $f^{\nu}_{\rm SF}=\varepsilon^2\,f^{\nu}_{(1)}+\varepsilon^3\,f^{\nu}_{(2)}+\cdots$ further reveals the condition that $|\beta |:=\varepsilon \left| f^{t}_{(2)}/f^{t}_{(1)}\right| \ll 1$.  We show in Fig.~\ref{expand} $\varepsilon^{-1}\beta$ as a function of $r$. Clearly this condition is met everywhere for our entire mass ratio range. At large distances the post Newtonian expression is $$\varepsilon^{-1}\beta=-\frac{35}{12}\frac{M}{r}+\frac{30,523}{4,032}\left(\frac{M}{r}\right)^2-\frac{101}{8}\pi\left(\frac{M}{r}\right)^{5/2}
+\cdots$$

\begin{figure}
 \includegraphics[width=7.5cm]{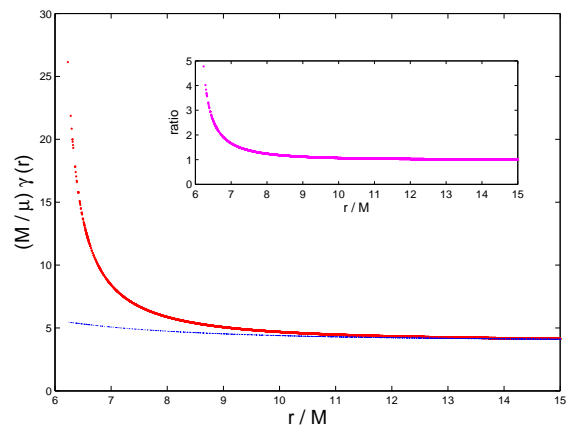}
\caption{The quantity $\varepsilon^{-1}\gamma$ as a function of $r/M$. The thick (red) curve is the curve based on the fully relativistic Barack--Sago results, and the thin dotted (blue) curve represents the first three terms in the post Newtonian expansion. The insert shows their ratio.}
\label{gamma}
\end{figure}

A perturbative expansion of SF driven quasi-circular Schwarzschild orbits was done in Ref.~\cite{burko-2003}. Specifically, the radial velocity was expanded in powers of $\varepsilon$ according to $v=\varepsilon v_{(1)}+\varepsilon^2v_{(2)}+\cdots$. Specifically, we define $\gamma:=\varepsilon v_{(2)}/v_{(1)}$. The results of \cite{burko-2003} imply that typical expressions for the radial velocity terms are such that the condition that $\gamma\ll 1$ can be given by  
\begin{equation}\label{gamma_cond}
2\,\varepsilon\,\left(\frac{r}{M}\right)^2\,\frac{r-3M}{r-6M}\,f_{r}^{(1)}\ll 1\, .
\end{equation}
The scaling of $f_r^{(1)}$ with $\varepsilon^0$ implies that $\gamma\sim O(\varepsilon)$. We therefore plot  $\varepsilon^{-1}\gamma$ as a function of $r$ in Fig.~\ref{gamma}. The far field behavior of $\gamma$ can be described by the post Newtonian expansion, specifically
$$\varepsilon^{-1}\gamma=4-2\left(\frac{M}{r}\right)+\frac{213}{5}\left(\frac{M}{r}\right)^2+\cdots$$ The condition (\ref{gamma_cond}) and Fig.~\ref{gamma} allow us, for each value of the mass ratio $\varepsilon$, to determine the domain of validity of our perturbative expansion. 

We consider the perturbative expansion again below in Section \ref{revisit}. 

\subsection{The waveforms}\label{wf}

We produce three waveforms: first, the waveforms for the case that only first order dissipative terms are included, and all higher order terms are excluded (``the energy balance waveform''; dephasing to $O(\varepsilon^{-1})$ (hereafter WF--I); the waveforms for the case that in addition to the former the first order conservative term at $O(\varepsilon^0)$ are included (WF--II); and the waveforms that in addition to the former also the second order dissipative term, also at $O(\varepsilon^{0})$ (WF--III). The three different cases are summarized in Table \ref{sum}. 

\begin{table}[htdp] 
\caption{Summary of the orbits and corresponding waveforms. The columns describe the effect for the phase of the gravitation wave, and the rows are the three types of waveforms that we consider. See the text for more detail. }
\begin{center}
\begin{tabular}{||c|c|c|c||}\hline\hline
 & Dissipative  & Conservative & Dissipative \\ 
 & at $O(\varepsilon^{-1})$ & at $O(\varepsilon^{0})$ & at $O(\varepsilon^{0})$ \\ \hline
WF--I & + & -  & -  \\ \hline
WF--II &+  & +  &  - \\ \hline
WF--III & + & + & + \\
\hline\hline
\end{tabular}\label{sum}
\end{center}
\label{default}
\end{table}%

\begin{figure}
 \includegraphics[width=7.5cm]{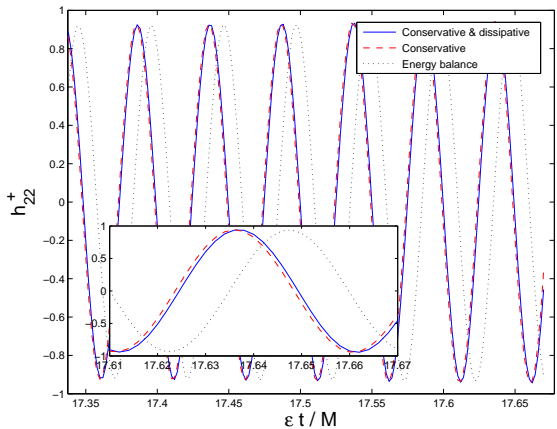}
\caption{The three waveforms for the mode $h_{22}^{\, +}$ for the mass ratio $\varepsilon=10^{-3}$ for a particle that starts at $r_0=8M$ and whose orbit decays down to the ISCO as function of $\varepsilon t/M$. The observer is positioned on the equatorial plane. 
The insert shows the last wavelength. We show WF--I with the dotted curve (black), WF--II with the red curve (dashed), and WF--III with the blue curve (solid). }
\label{wf001}
\end{figure}

\begin{figure}
 \includegraphics[width=7.5cm]{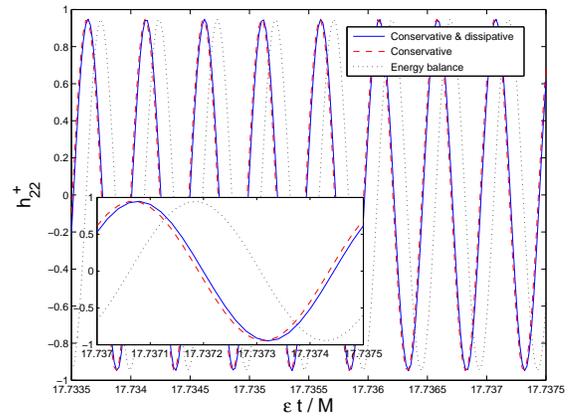}
\caption{Same as Fig.~\ref{wf001} for the mass ratio $\varepsilon=10^{-5}$.}
\label{wf00001}
\end{figure}

Figures \ref{wf001} and \ref{wf00001} show the $h_{22}^{+}$ mode of the waveform for particles of mass ratios $10^{-3}$ and $10^{-5}$, correspondingly, that start inspiraling from $r_0=8M$ down to the ISCO (in practice we let the particles get to $6.01M$) as functions of the time. In all cases the three waveforms are in phase at $t=0$, and the dephasing has not gone over $2\pi$ radians.

\begin{figure}
 \includegraphics[width=8.0cm]{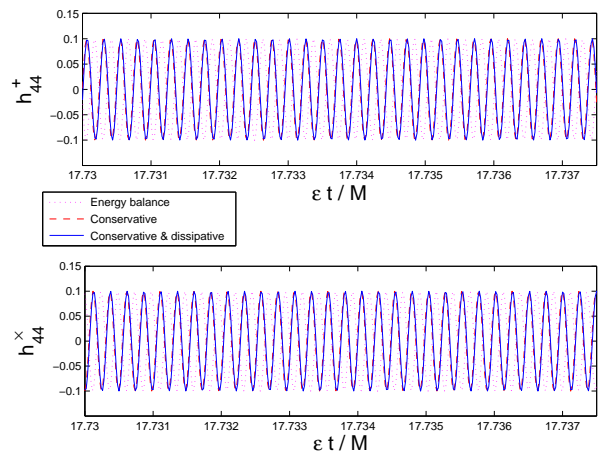}
\caption{Same as Fig.~\ref{wf001}  for the mass ratio $\varepsilon=10^{-5}$ for the modes $h_{44}^{+,\times}$.}
\label{wf_h44}
\end{figure}

Figure \ref{wf_h44} shows the waveforms for the modes $h_{44}^{+,\times}$ for the mass ratio $\varepsilon=10^{-5}$ for the same orbit as shown in Fig.~\ref{wf00001}. At $t=0$ all three waveforms are in phase for both polarization states. Other modes can be shown similarly. 

Figures \ref{wf001}, \ref{wf00001}, and \ref{wf_h44} suggest that the dephasing between WF--III and WF--II is substantially smaller than the dephasing between WF--II and WF--I. We study this conclusion quantitatively below in Section \ref{dephasing}. 

\subsection{The perturbative expansion revisited}\label{revisit}

The smallness of the effect raises the question of whether the second order dissipative effect is negligible. In Ref.~\cite{burko-2003} this question was raised, and specifically a condition for neglecting the second order dissipative effect was considered. This condition is given by
\begin{equation}
|\alpha |:=\frac{1}{2}\,\left(1-2\frac{M}{r}\right)\,\left(\frac{M}{r}\right)^2\,\frac{r-6M}{r-3M}\,   \left| \frac{f^t_{(2)}}{f^t_{(1)}\;f^r_{(1)}}\right| \ll 1\, .
\end{equation}
The scaling of the self force components with $\varepsilon$ implies that $\alpha$ is at $O(\varepsilon^0)$, i.e., independent of the mass ratio.  
At great distances $\alpha$ can be given by its post Newtonian expression as
\begin{equation}
\alpha=-\frac{35}{48}\left(\frac{M}{r}\right)+\frac{48,163}{16,128}\left(\frac{M}{r}\right)^2-\frac{101}{32}\pi\left(\frac{M}{r}\right)^{5/2}+\cdots
\end{equation}

\begin{figure}
 \includegraphics[width=7.5cm]{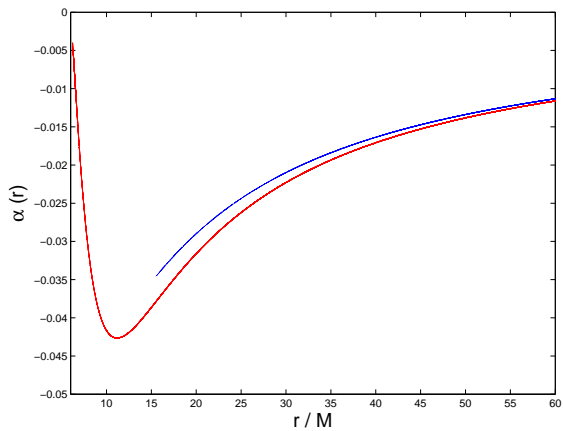}
\caption{The variable $\alpha$ as a function of distance. The solid (red) curve is the fully relativistic expression. The shorter dotted (blue) curve is the first post Newtonian term. }
\label{alpha}
\end{figure}

Figure \ref{alpha} shows $\alpha$ as a function of $r$. Both at great distances and very close to the ISCO $\alpha$ is very small, specifically $\alpha\to 0^-$ for either $r\to\infty$ or $r\to 6M$. The parameter $|\alpha¨|$ is maximal around $r\sim 11M$, and is never greater than $~0.043$. As $\alpha$ is independent of the mass ratio, we conclude that justification for neglecting the second order dissipative effect can be found only for orbits very close to the ISCO (or alternatively for orbits that are only very far from it).

\section{The dephasing}\label{dephasing}

\begin{figure}[h]
 \includegraphics[width=7.5cm]{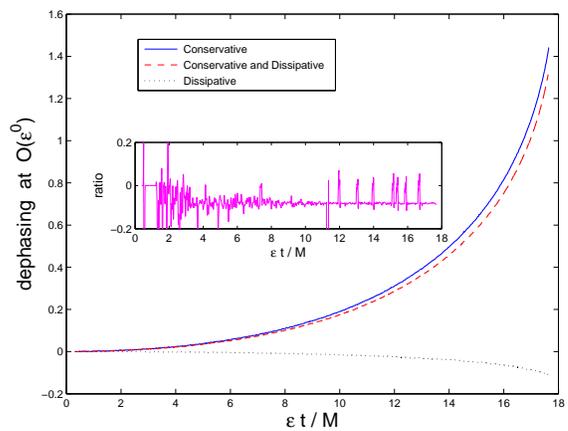}
\caption{The dephasing at $O(\varepsilon^0)$ between the different $h_{22}^+$ waveforms as functions of $\varepsilon t/M$. When plotted against this variable, the figure is independent of the mass ratio $\varepsilon$. The solid (blue) curve shows the dephasing between WF--I and WF--II, the dashed (red) curve shows the dephasing between WF--I and WF--II, and the dotted (black) curve shows the dephasing between WF--II and WF--III. The insert shows the ratio of the dotted to the solid curves. 
}
\label{deph}
\end{figure}

\begin{figure}[h]
 \includegraphics[width=7.5cm]{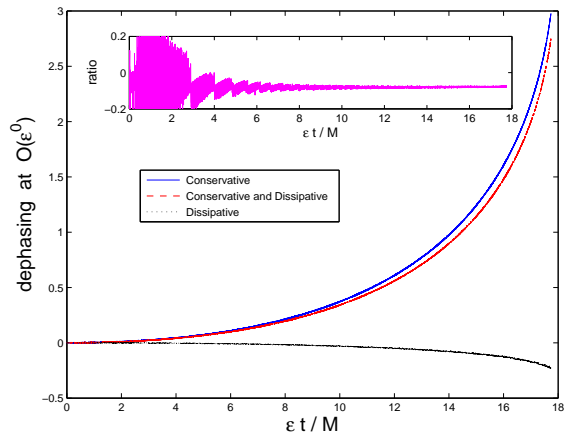}
\caption{Same as Fig.~\ref{deph} for the $h_{44}^+$ mode. Notice that the dephasing itself is greater than for the $h_{22}^+$ mode, as can be expected from the shorter wavelength of the $h_{44}^+$ mode.}
\label{deph44}
\end{figure}

We study next the dephasing between the three waveforms that we presented above in Sec.~\ref{wf}. We present in Fig.~\ref{deph} the relative dephasings between the waveforms as functions of the time for the $h_{22}^+$ mode, and in Fig.~\ref{deph44} we plot the same for the $h_{44}^+$ mode. The dephasing of  WF--I and WF--II was presented in Paper I. For ease of comparison, it is reproduced in Figs.~\ref{deph},\ref{deph44}. We also show in Figs.~\ref{deph},\ref{deph44} the dephasing of WF--I and WF--III, and the dephasing of WF--I and the waveform by adding only the $O(\epsilon^0)$ contribution of the second--order dissipative effect to the first order dissipative case. We also show in Figs.~\ref{deph},\ref{deph44} the ratio of the latter the the second case. Except for the expected numerical noise, this ratio is at a nearly constant level of $8\%$ throughout the evolution. All the curves in Figs.~\ref{deph},\ref{deph44} are independent of $\varepsilon$, and the curves for the various values of the mass ratio that we considered (in the range $10^{-5}$--$10^{-2}$) overlap each other in this figure.

Notice that the contribution of the the second order dissipative effect is to reduce the overall dephasing of the waveform at $O(\varepsilon)$ with the waveform at $O(\varepsilon^{-1})$. This result is indeed expected as the sign of $f^t_{(2)}$ is opposite to the sign of $f^t_{(1)}$.

The contribution of the dephasing at $O(\varepsilon^0)$ is therefore dominated by the first order conservative effect. The contribution of the second order dissipative effect is to decrease the total dephasing, by an amount of about $8\%$ of the total $O(\varepsilon^0)$ effect. We therefore suggest that for gravitational wave detection the second order dissipative effect is not very important, although it may be so for accurate gravitational wave astronomy, e.g., for accurate parameter estimation. Neglecting this effect would result in an inconsistency in the determination of the mass ratio, specifically in increasing the error bars of the measurement. Figures \ref{deph} and \ref{deph44} are  our main result in the present Paper.

\begin{figure}
 \includegraphics[width=7.5cm]{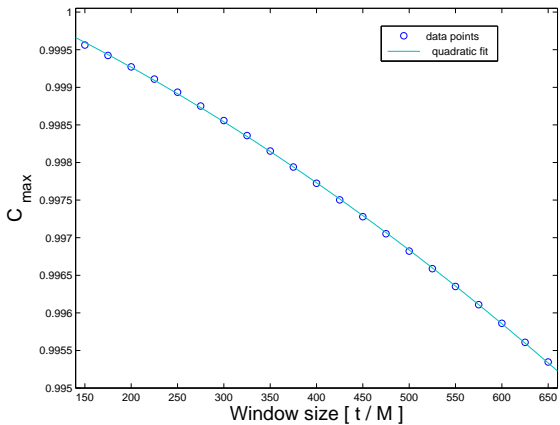}
\caption{Overlap integral for the full waveform WF--III with a sliding window taken from WF--II as a function of the window's width,  for the mass ratio $\varepsilon=10^{-5}$. The circles represent the data points, and the solid curve is a quadratic fit.}
\label{window}
\end{figure}

The relatively small contribution to the total dephasing of the second order dissipative effect is further manifest in the overlap integral between the full waveform at $O(\varepsilon^0)$, i.e., WF--III, and an integration window selected from the waveform that includes at $O(\varepsilon^0)$ only the first order conservative effect (WF--II). We select the window from the end of the waveform, and find the overlap integral as a function of the width of the integration window. In Fig.~\ref{window} we plot the overlap integral as a function of the window width. Comparing Fig.~\ref{window} with Fig.~12 of Paper I, we see that the contribution of the second order dissipative effect to the total overlap integral is at the order of $10^{-3}$. 

\section{The gauge invariant plot}\label{results}

We showed in Paper I that when the component $u^t$ of the four-velocity is plotted against the angular frequency $\Omega$ for WF--I and WF--II, the two curves overlap. That is, when the orbit of the small mass $\mu$ is represented on this gauge invariant plot, the curve is unchanged (or at least, is changed only very little). However, we showed in Paper I that the speed by which the data point representing the system moves along the curve on the gauge invariant curve depends on whether one includes or excludes the first order conservative effect. Specifically, with the inclusion of the first order conservative effect the system moves faster along the gauge invariant curve compared with the energy balance case. 

\begin{figure}
 \includegraphics[width=7.5cm]{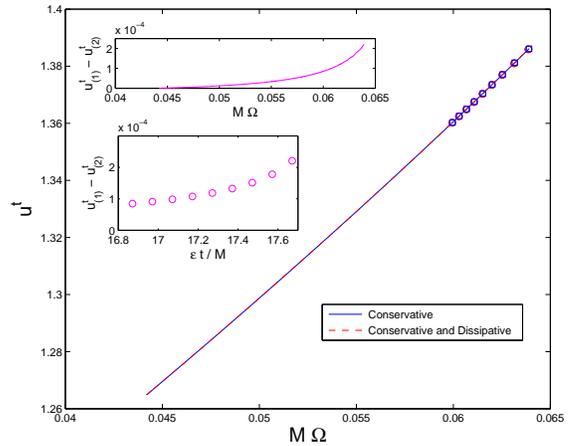}
\caption{The gauge invariant plot for the mass ratio $\varepsilon=10^{-5}$, showing $u^t$ as a function of $\Omega M$ for two cases: for WF--II (solid blue curve, $\Box$) and for WF--III (dashed red curve, $\circ$). The two curves are indistinguishable on this figure, and are independent of $\varepsilon$. We also show equally spaced (in time, with increments of $10,000M$) data points. The data points corresponding with WF--III advance slower along the curve than those corresponding with WF--II. The upper insert shows the different between the WF--II case (denoted here $u^t_{(1)}$) and the WF--III case (denoted here $u^t_{(2)}$) as a function of $\Omega M$, and the lower insert shows the same as a function of $\varepsilon t/M$ (not gauge invariant!) for the same data points as in the main figure.}
\label{gi}
\end{figure}

In Fig.~\ref{gi} we present the gauge invariant curves for WF--II and for WF--III. On the scale of Fig.~\ref{gi} the two curves completely overlap. However, there is a small difference between them, at order $10^{-4}$. However, a data point describing a system evolving according the WF--III moves slower along the curve than a data point describing a system evolving according to WF--II. The effect of the second order dissipative term is to retard the orbital evolution. The effect, however, is so small, that it is hard to see how the gauge invariant plot could be used to distinguish between templates that include or exclude the second order dissipative term. 

\section*{Ackowledgments} 

We thank L.~Barack for discussions. L.M.B.~ acknowledges support from NSF grants PHY--1249302 and EDU--1307148. 
G.K.~acknowledges research support from NSF grants PHY--1016906, CNS--0959382 and PHY--1135664, and from the US Air Force Grant Nos.~FA9550--10--1--0354 and 10--RI--CRADA--09. All computations were performed on the NSF XSEDE {\em Keeneland} GPU supercomputer under XRAC allocation PHY--120037.

\begin{appendix}
\section{The self force expressions}\label{PN}

We expand the self force as $f^{\nu}_{\rm SF}=\varepsilon^2\,f^{\nu}_{(1)}+\varepsilon^3\,f^{\nu}_{(2)}+\cdots$. The expressions we use for the self force are given by (see Paper I for more detail):

 \begin{eqnarray}
f^t_{(1)\;r\le 8\,M}&=&-
\frac{1}{\sqrt{1-3x}\left(1-2x \right)} \; x^5\;
\Bigg[a^-_0\nonumber\\
&+&a^-_1\,x+a^-_2\, x^2+a^-_3\, x^3 + \cdots \Bigg]
\end{eqnarray}
\begin{eqnarray}\label{pn}
f^t_{(1)\;r\ge 8\,M}&=&-
\frac{32}{5}\,\frac{1}{\sqrt{1-3x}\left(1-2x \right)} \; x^5\;\\
&\times&\;
\left[ {\rm PN}_{5.5}+\left(a^+_6+a^+_{6L}\,\ln x \right) x^6 + \cdots \right]\nonumber
\end{eqnarray}
\begin{eqnarray}
f^r_{(1)\; r\le 8\,M}&=&
\bigg(1-2x\bigg)\; x^2
\nonumber\\
&\times&\;
\Bigg[b^-_0+b^-_1\,\bigg(1-6x\bigg)+b^-_2\,\bigg(1-6x \bigg)^2\nonumber\\
&+&b^-_3\,\bigg(1-6x \bigg)^3 + \cdots\Bigg]
\end{eqnarray}
\begin{eqnarray}
f^r_{(1)\;r\ge 8\,M}&=& x^2
\;
\left( b^+_0
+ b^+_1\, x+b^+_2\, x^2
+b^+_3\, x^3 + \cdots\right)
\end{eqnarray}

\begin{eqnarray}
f^t_{(2)}=&-& \frac{32}{5} \frac{1}{\sqrt{1-3x}\left(1-2x \right)}
\, x^6\,\left[ - \frac{35}{12}+\frac{9,271}{504}\, x \right. \nonumber \\
&-&\left. \frac{583}{24}\,\pi\, x^{3/2} +\left( -\frac{134,543}{7,776}+\frac{41}{48}\,\pi^2\right)\,x^2\right. \nonumber \\
&+&\left. \frac{214,745}{1,728}\, x^{5/2}+ \cdots \right]
\end{eqnarray}
where $x:=(M\Omega)^{2/3}$ is the dimensionless orbital frequency. Here, 
\begin{eqnarray*}
{\rm PN}_{5.5}&=& 1 - \frac{1,247}{336} x + 4 \pi x^{\frac{3}{2}} - 
 \frac{44,711}{9,072}x^2 \\
 &-& \frac{8,191}{672}\ \pi x^{\frac{5}{2}}+  
\Bigg( \frac{6,643,739,519}{69,854,400} 
 - \frac{1,712}{105}\,  \gamma  
+  \frac{16}{3}\  \pi^2\\ 
&-& \frac{3,424}{105}\  \ln 2 
 -\frac{856}{105} \,\ln x \Bigg)x^3 
   - \frac{16,285}{504} \pi x^{7/2} \\
 &+& \Bigg(-\frac{323,105,549,467}{3,178,375,200}+ 
\frac{232,597}{4,410}\, \gamma - \frac{1,369}{126}\  \pi^2 \\
&+& \frac{39,931}{294}\  \ln 2 - \frac{47,385}{1,568}\  \ln 3  +  
  \frac{232,597}{8,820}\ \ln x\Bigg) x^4\\
  &+& \pi\, \Bigg(\frac{265,978,667,519}{745,113,600} - \frac{6,848}{105}\, \gamma
-  \frac{13,696}{105}\, \ln 2 \\
&-&  \frac{3,424}{105}\ \ln x \Bigg)x^{9/2}
+\ \Bigg(-\frac{2,500,861,660,823,683}{2,831,932,303,200} \\
&+& \frac{916,628,467}{7,858,620}\ \gamma
- \frac{424,223}{6,804}\,\pi^2 
- \frac{83,217,611}{1,122,660}\ \ln 2\\
&+& \frac{47,385}{196}\  \ln 3
+  \frac{916,628,467}{15,717,240}\ \ln x\Bigg) x^{5} \\
   %%%%%%%%%%%%%%
&+&\pi\, \Bigg(\frac{8,399,309,750,401}{101,708,006,400}
+ \frac{177,293}{1,176}\ \gamma
\\
&+&   \frac{8,521,283}{17,640} \ \ \ln 2-\frac{142,155}{784} \ \ \ln 3 \\
&+&  \frac{177,293 }{2,352}\  \ln x\Bigg)x^{11/2}\, ,
\end{eqnarray*}
and the expansion parameters are given in Table \ref{default}. 
\begin{table}[htdp]
\caption{The fit parameters for the self force. These parameters reproduce the accuracy of \cite{barack-sago} to all significant figures for all data points.}
\begin{center}
\begin{tabular}{|c|c||c|c||c|c||c|c||}\hline\hline
$a^-_0$ & 4.57583 & $a^+_6$ & 331.525 & $b^-_0$ & 1.32120 & $b^+_0$ & 1.999991\\
\hline
$a^-_1$ & 31.8117 & $a^+_{6L}$ & -2081.57 & $b^-_1$ & 1.2391 & $b^+_1$ & -6.9969\\
\hline
$a^-_2$ & -267.250 & & & $b^-_2$ & -1.297 & $b^+_2$ & 6.29\\
\hline
$a^-_3$ & 1049.27 & & & $b^-_3$ & 1.07 & $b^+_3$ & -24.6\\
\hline\hline
\end{tabular}
\end{center}
\label{default}
\end{table}%

\end{appendix}

\end{document}